\documentclass[11pt]{article}

\usepackage{graphicx}
\usepackage{hyperref}

\begin{document}

\baselineskip=17pt

%%%% *****************************************************************
%%%% *************    Text stat'i     ********************************
%%%% *****************************************************************
\newpage
\pagenumbering{arabic}

\begin{center}
\Large \bf High energy particles with negative and positive energies in
the vicinity of black holes
\end{center}

\begin{center}
\bf
A. A. Grib${}^{1,2,3,a}$\footnote{${}^{a}$E-mail:\, andrei\_grib@mail.ru},
Yu. V. Pavlov${}^{1,2,4,b}$\footnote{${}^{b}$E-mail:\, yuri.pavlov@mail.ru}
\end{center}

\begin{center}
  ${}^{1}$\,A. Friedmann Laboratory for Theoretical Physics,
Saint Petersburg, Russia;\\
  ${}^{2}$\,Copernicus Center for Interdisciplinary Studies,
Krak\'{o}w, Poland, EU;\\
  ${}^{3}$\,Theoretical Physics and Astronomy Department,
The Herzen  University\\
Moika 48, Saint Petersburg 191186, Russia;\\
  ${}^{4}$\,Institute of Problems in Mechanical Engineering,
Russian Academy of Sciences\\
Bol'shoy pr. 61, V.O., Saint Petersburg 199178, Russia
\end{center}

\begin{abstract}
    It is shown that the energy in the centre of mass frame of two colliding
particles in free fall at any point of the ergosphere of the rotating black
hole can grow without limit for fixed energy values of particles on infinity.
    The effect takes place for large negative values of the angular momentum
of one of the particles.
    It occurs that the geodesics with negative energy in equatorial plane of
rotating black holes cannot originate or terminate inside the ergosphere.
    Their length is always finite and this leads to conclusion that they
must originate and terminate inside the gravitational radius of the ergosphere.
    The energy in the centre of mass frame of one particle falling into the
gravitational radius and the other arriving from the area inside it is growing
without limit on the horizon.
\end{abstract}

{\small
{\bf Key words:} \, black holes, Kerr metric, geodesics, negative energy,
particle collisions.

{\bf PACS numbers:} \, 04.70.-s, 04.70.Bw, 97.60.Lf

%%\pacs{04.70.-s}{Physics of black holes}
%%\pacs{04.70.Bw}{Classical black holes}
%%\pacs{97.60.Lf}{Black holes}
}

\vspace{2mm}
%%%% *****************************************************************
{\centering \section{\large Introduction}}
\label{secIntr}

    The main part of the paper is a a review of our results obtained
in~\cite{GribPavlov2012}--\cite{GribPavlov2013a}.
    In the end we discuss some new features of particles with negative energies
concerning their origination and collisions with ordinary particles.

    Recently many papers were
published~\cite{BanadosSilkWest09}--\cite{Zaslavskii12b} in which some
specific properties of collisions of particles close to the horizon of
the rotating black holes were discussed.
    In~\cite{BanadosSilkWest09} the effect of unlimited growing energy of
colliding two particles in the centre of mass frame for critical Kerr
rotating black holes (call it BSW effect) was discovered.
    In our papers~\cite{GribPavlov2010}--\cite{GribPavlov2011b}
the same effect was found for noncritical black holes when multiple
collisions took place.

    All this shows that close to the horizon there is natural supercollider
of particles with the energies up to Planck's scale.
    However its location close to the horizon makes difficult due to the large
red shift get some observed effects outside of the ergosphere when particles
go from the ``black hole supercollider'' outside.
    Here we shall discuss some other effect valid at any point of
the ergosphere.
    The energy in the centre of mass frame can take large values for large
negative values of the angular momentum projection on the rotation axis of
the black hole.
    The problem is how to get this large (in absolute value) negative values.
    That is why first we shall obtain limitations on the values of
the projection of angular momentum outside ergosphere for particles falling
into the black hole and inside the ergosphere.
    It occurs that inside the ergosphere of black hole there is no limit for
negative values of the momentum and arbitrary large energy in the centre
of mass frame is possible for large angular momentum of the two colliding
particles.
    In a sense the effect is similar to the BSW-effect.
    One of the particle with large negative angular momentum can be called
``critical'', the other particle ``ordinary''.

    Also in this paper we analyse geodesics for particles with negative energies
for rotating black holes.
    Kerr's metric~\cite{Kerr63}, predicts differently from the Schwarzschild
metric  existence of the special region outside the horizon of the black hole
called ergosphere.
    In the egosphere elementary particles must rotate together with the black
hole.
    The new feature of ergosphere is the existence of geodesics with negative
relative to infinity energy.
    Existence of such geodesics leads to the possibility of extraction of
energy from the rotating black holes due to the
Penrose process~\cite{Penrose69}
    so one can call these geodesics the Penrose geodesics.
    However in spite of the more than 40 years passing after the discovery of
the Penrose effect there is still no information about the full picture of the
Penrose geodesics especially about their origin.
    Here we shall investigate the problem of properties of such geodesics.
    Particle can arrive on such trajectory in the result of collisions or
decays in the ergosphere.
    But the world line of the geodesic in geodesically complete space-time
must originate or terminate either in singularity or in
the infinity~\cite{HawkingEllis}.
    Note that here one means infinity in space-time so that it can be infinity
in time for finite value of the space distances.
    The problem is to find where originate and terminate Penrose geodesics and
this is the subject of the paper.
    It will be shown that the length of the Penrose geodesics is always finite
inside the ergosphere so that the only possibility of their origination and
termination is outside of the ergosphere.
    But the geodesics for particles with negative energy don't exist in the
external space out of ergosphere so that one comes to the conclusion that they
originate and terminate inside the gravitational radius of the black hole.
    The fact that they terminate inside the horizon is well
known~\cite{Contopoulos84}.
    So our interest in this paper is to find the place of their origination.
    These geodesics occur to be ``white hole'' geodesics originating at the
singularity arriving to ergosphere and then returning back to the horizon
and singularity.
    A new feature is that if particle with negative energy collides with
ordinary particle the energy in the centre of mass frame is growing without
limit on the horizon.

    The system of units $G=c=1$ is used in the paper.

\vspace{9mm}
%%%% *****************************************************************
{\centering \section{General formulas for the energy of particles
close to the black hole}
\label{sec2}}

    The Kerr's metric of the rotating black hole~\cite{Kerr63} in
Boyer--Lindquist coordinates~\cite{BoyerLindquist67} has the form:
    \begin{equation}
d s^2 = d t^2 -
\frac{2 M r}{\rho^2} \, ( d t - a \sin^2 \! \theta\, d \varphi )^2
-\, \rho^2 \left( \frac{d r^2}{\Delta} + d \theta^2 \right)
- (r^2 + a^2) \sin^2 \! \theta\, d \varphi^2,
\label{Kerr}
\end{equation}
    where
    \begin{equation} \label{Delta}
\rho^2 = r^2 + a^2 \cos^2 \! \theta, \ \ \ \ \
\Delta = r^2 - 2 M r + a^2,
\end{equation}
    $M$ is the mass of the black hole, $ aM $ its angular momentum.
    The rotation axis direction corresponds to $\theta =0$, i.e. $a \ge 0$.
    The event horizon of the Kerr's black hole corresponds to
    \begin{equation}
r = r_H \equiv M + \sqrt{M^2 - a^2} .
\label{Hor}
\end{equation}
    For Cauchy horizon one has
    \begin{equation}
r = r_C \equiv M - \sqrt{M^2 - a^2} .
\label{HorC}
\end{equation}
    The surface of the static limit is defined by
    \begin{equation}
r = r_1(\theta) \equiv M + \sqrt{M^2 - a^2 \cos^2 \theta} .
\label{Lst}
\end{equation}
    In case $ a \le M $ the region of space-time between the static limit
and event horizon is called ergosphere.

    For geodesics in Kerr's metric~(\ref{Kerr}) one obtains
(see \cite{Chandrasekhar}, Sec.~62 or \cite{NovikovFrolov}, Sec.~3.4.1)
    \begin{equation} \label{geodKerr1}
\rho^2 \frac{d t}{d \lambda } = -a \left( a E \sin^2 \! \theta - J \right)
+ \frac{r^2 + a^2}{\Delta}\, P,
\end{equation}
    \begin{equation}
\rho^2 \frac{d \varphi}{d \lambda } =
- \left( a E - \frac{J}{\sin^2 \! \theta} \right) + \frac{a P}{\Delta} ,
\label{geodKerr2}
\end{equation}
    \begin{equation} \label{geodKerr3}
\rho^2 \frac{d r}{d \lambda} = \sigma_r \sqrt{R}, \ \ \ \
\rho^2 \frac{d \theta}{d \lambda} = \sigma_\theta \sqrt{\Theta},
\end{equation}
    \begin{equation} \label{geodP}
P = \left( r^2 + a^2 \right) E - a J,
\end{equation}
    \begin{equation} \label{geodR}
R = P^2 - \Delta [ m^2 r^2 + (J- a E)^2 + Q],
\end{equation}
    \begin{equation} \label{geodTh}
\Theta = Q - \cos^2 \! \theta \left[ a^2 ( m^2 - E^2) +
\frac{J^2}{\sin^2 \! \theta} \right].
\end{equation}
    Here $E$ is conserved energy (relative to infinity)
of the probe particle,
$J$ is conserved angular momentum projection on the rotation axis
of the black hole,
$m$ is the rest mass of the probe particle, for particles with nonzero
rest mass $\lambda = \tau /m $,
where $\tau$ is the proper time for massive particle,
$Q$ is the Carter's constant.
    The constants $\sigma_{r}, \sigma_{\theta}$ in formulas~(\ref{geodKerr3})
are equal to $\pm 1$ and are defined by the direction
of particle movement in coordinates $r$, $\theta$.
    For massless particles one must take $m = 0$
in~(\ref{geodR}), (\ref{geodTh}).

%%%%%%%%%%%%%%%%%%%%%%%%%%%%%%%%%%%%%%%%%%%%%%%%%%
    The permitted region for particle movement outside the event horizon
is defined by conditions
    \begin{equation} \label{ThB0}
R \ge 0, \ \ \ \ \ \Theta \ge 0, \ \ \ \ \
\frac{d t}{d \lambda} \ge 0 .
\end{equation}
    The last inequality forbids movement ``back in time''~\cite{Wald}.
    Let us find limitations for the particle angular momentum from the
conditions~(\ref{ThB0}) at the point $(r, \theta)$,
taking the fixed values of $\Theta$~\cite{GribPavlov2013}.

    Outside the ergosphere $ r^2 -2 r M +a^2 \cos^2 \! \theta >0 $ one obtains
    \begin{equation} \label{EvErg}
E \ge \frac{1}{\rho^2} \sqrt{(m^2 \rho^2 + \Theta)
(r^2 -2 r M +a^2 \cos^2 \! \theta)}, \ \ \ \ \
    %% \end{equation}    \begin{equation} \label{JvErg}
J \in \left[ J_{-}, \ J_{+} \right],
\end{equation}
    \begin{equation}
J_{\pm} = \frac{\sin \theta}{r^2 -2 r M +a^2 \cos^2 \! \theta}
\biggl[ - 2 r M a E \sin \theta
\pm \sqrt{ \Delta \left( \rho^4 E^2 \!-\! (m^2 \rho^2 \!+\! \Theta)
(r^2 \!-\! 2 r M \!+\! a^2 \cos^2 \! \theta) \right)} \biggr] .
\label{Jpm}
\end{equation}

    On the boundary of ergosphere
    \begin{equation} \label{rEgErg}
r = r_1(\theta) \ \ \ \Rightarrow \ \ \ E \ge 0, \ \ \ \
J \le E \left[ \frac{M r_1(\theta) }{a} + a \sin^2 \! \theta \left(\!
1 - \frac{m^2}{2 E^2} - \frac{\Theta}{4 M r_1(\theta) E^2} \!\right)
\! \right]\!.
\end{equation}

    Inside ergosphere
    \begin{equation} \label{lHmdd}
r_H < r < r_1(\theta) \ \ \ \Rightarrow \ \ \
(r^2 -2 r M +a^2 \cos^2 \! \theta) <0 ,
\end{equation}
    \begin{equation}
J \le J_{-}(r,\theta) = \frac{- \sin \theta}{
r^2 \!-\! 2 r M \!+\! a^2 \cos^2 \! \theta} \biggl[ 2 r M a E \sin \theta
- \sqrt{ \Delta \left( \rho^4 E^2 \!-\! (m^2 \rho^2 + \Theta)
(r^2 \!-\! 2 r M \!+\! a^2 \cos^2 \! \theta) \right)} \biggr].
\label{JmErg}
\end{equation}
    So it is only inside the ergosphere that the energy $E$ of the particle
relative to infinity can be negative.
    From~(\ref{JmErg}) one can see that for negative energy~$E$ of the particle
in ergosphere its angular momentum projection on the rotation axis of the
black hole must be also negative.

    For negative values of the energy $E$ the function $ J_{-}(r,\theta) $
is decreasing with growing $r$ in ergosphere, so that
    \begin{equation} \label{JgEHf}
\theta \ne 0 , \pi , \ \ \ r \to r_1(\theta) \ \ \Rightarrow \ \
J_{-}(r,\theta) \to - \infty .
\end{equation}
    So in order to come to the upper frontier of the ergosphere particle
with negative energy one must have infinitely large in absolute value negative
angular momentum.

    In the limit $r \to r_H$ from~(\ref{JmErg}) (for $\theta \ne 0, \pi$)
one obtains
    \begin{equation} \label{JgEH}
J \le J_H = \frac{ 2 M r_H E}{a} .
\end{equation}
    So $J_H$ is the maximal value of the angular momentum of the particle with
the energy~$E$ close to the gravitational radius.

\vspace{9mm}
%%%% *****************************************************************
{\centering \section{The energy of particles collision close to the black hole}
\label{secCollision}}

    One can find the energy in the centre of mass frame of two colliding
particles $E_{\rm c.m.}$ with rest masses~$m_1$, $m_2$ taking the square of
    \begin{equation} \label{SCM}
\left( E_{\rm c.m.}, 0\,,0\,,0\, \right) = p^{\,i}_{(1)} + p^{\,i}_{(2)},
\end{equation}
    where $p^{\,i}_{(n)}$ are 4-momenta of particles $(n=1,2)$.
    Due to $p^{\,i}_{(n)} p_{(n)i}= m_n^2$ one has
    \begin{equation} \label{SCM2af}
E_{\rm c.m.}^{\,2} = m_1^2 + m_2^2 + 2 p^{\,i}_{(1)} p_{(2)i} .
\end{equation}
    Note that the energy of collisions of particles in the centre of mass
frame is always positive (while the energy of one particle due to Penrose
effect~\cite{Penrose69} can be negative!) and satisfies the condition
    \begin{equation} \label{Eb0}
E_{\rm c.m.} \ge m_1 + m_2.
\end{equation}
    This follows from the fact that the colliding particles move one
towards another with some velocities.

    It is important to note that $E_{\rm c.m.}$ for two colliding particles
is not a conserved value differently from energies of particles (relative
to infinity) $E_1$,  $E_2$.

    For the free falling particles with energies $E_1$,  $E_2$ and angular
momentum projections $J_1, J_2$ from~(\ref{geodKerr1})--(\ref{geodR})
one obtains~\cite{HaradaKimura11}:
    \begin{eqnarray}
E_{\rm c.m.}^{\,2} = m_1^2 + m_2^2 + \hspace{57mm}
\nonumber \\[7pt]
+\, \frac{2}{\rho^2} \biggl[ \, \frac{P_1 P_2 -
\sigma_{1 r} \sqrt{R_1} \, \sigma_{2 r} \sqrt{R_2}}{\Delta}
- \frac{ (J_1 - a  E_1 \sin^2 \! \theta) (J_2 - a  E_2 \sin^2 \! \theta)}
{\sin^2 \! \theta}
 - \sigma_{1 \theta} \sqrt{\Theta_1} \,
\sigma_{2 \theta} \sqrt{\Theta_2} \, \biggr].
\label{KerrL1L2}
\end{eqnarray}
    The big values of the collision energy can occur near the event horizon
if one of the particles has the ``critical'' angular moment $J_H$:
    \begin{eqnarray}
E_{\rm c.m.}^{\,2}(r \to r_H) = \frac{ (J_{1H} J_2 - J_{2H} J_1)^2}
{4 M^2 (J_{1H} - J_1) (J_{2H} - J_2)}
+ m_1^2 \left[1+ \frac{J_{2H} \!- J_2}{J_{1H} \!- J_1}\right] +
m_2^2 \left[ 1+ \frac{J_{1H} \!- J_1}{J_{2H} \!- J_2}\right]
\label{GrPvPi}
\end{eqnarray}
(see Eq.~16 in~\cite{GribPavlovPiattella2012}).
    This is the BSW-effect.

    The energy in the centre of mass frame can be written
through the relative velocity~$ v_{\rm rel}$ of particles at the moment
of collision~\cite{Zaslavskii11}, \cite{GribPavlovPiattella2012}:
    \begin{equation} \label{Relsk03}
E_{\rm c.m.}^{\,2} = m_1^2 + m_2^2 +
\frac{2 m_1 m_2}{\sqrt{1 \!- v_{\rm rel}^2}}
\end{equation}
    and the nonlimited growth of the collision energy in the centre of mass
frame occurs due to growth of the relative velocity to the velocity of
light~\cite{Zaslavskii11}.

    Let's consider another opportunity for big values of collision energy.
    As we can see from~(\ref{JmErg}),
on the boundary and inside ergosphere there exist geodesics on
which particle with fixed energy can have arbitrary large in absolute value
negative angular momentum projection.
    Let us find the asymptotic of~(\ref{KerrL1L2}) for $J_2 \to -\infty$ and
some fixed value $r$ in ergosphere supposing the value of Carter's
constant $Q_2$ to be such that~(\ref{ThB0}) is valid and $\Theta_2 \ll J_2^2$.
    Then from~(\ref{KerrL1L2}) one obtains
    \begin{eqnarray}
E_{\rm c.m.}^{\,2} \approx \frac{- 2 J_2}{\rho^2 \Delta } \,
\biggl[ \frac{J_1}{\sin^2 \! \theta}
\left( r^2 \!-\! 2 r M \!+\! a^2 \cos^2 \! \theta \right)
+ 2 r M a E_1
- \frac{\sigma_{1r} \sigma_{2r} \sqrt{R_1}}{\sin \theta}
\sqrt{-(r^2 \!-\! 2 r M \!+\! a^2 \cos^2 \! \theta) } \biggr] .
\label{KerrJB}
\end{eqnarray}
    This asymptotic formula is valid for all possible $E_1$, $J_1$
(see~(\ref{JmErg})) for $r_H < r < r_{1}(\theta)$ and for
$E_1>0$ and $J_1$ satisfying~(\ref{rEgErg}) for $r=r_{1}(\theta)$.
    The poles $\theta = 0, \pi$ are not considered here because the points
on surface of static limit are on the event horizon.

    Note that expression in brackets in~(\ref{KerrJB}) is positive
in ergosphere.
    This is evident for $r=r_{1}(\theta)$ and follows from
limitations~(\ref{JmErg}) for $r_H < r < r_{1}(\theta)$,
and inside ergosphere~(\ref{KerrJB}) can be written as
    \begin{eqnarray}
E_{\rm c.m.}^{\,2} \approx J_2 \frac{r^2 -2 r M +a^2 \cos^2 \! \theta}
{\rho^2 \Delta \sin^2 \! \theta}
\left( \sigma_{1r} \sqrt{J_{1 +}- J_1} -
\sigma_{2r} \sqrt{J_{1 -}- J_1} \right)^2.
\label{KerrJBner}
\end{eqnarray}

    So from~(\ref{KerrJB}) one comes to the conclusion that
{\it when particles fall on the rotating black hole collisions with arbitrarily
high energy in the centre of mass frame are possible at any point of
the ergosphere if $J_2 \to -\infty$ and the energies $E_1, E_2$ are fixed}.
    The energy of collision in the centre of mass frame is growing
proportionally to $\sqrt{|J_2|}$.

    Note that large negative values of the angular momentum projection are
forbidden for fixed values of energy of particle out of the ergosphere.
    So particle which is nonrelativistic on space infinity ($E=m$)
can arrive to the horizon of the black hole if its angular momentum projection
is located in the interval
    \begin{equation}
-2 m M \left[ 1 + \sqrt{1+ \frac{a}{M}}\, \right] \le J \le 2 m M
\left[ 1 + \sqrt{1 - \frac{a}{M}}\, \right].
\label{KerrEM}
\end{equation}
    The left boundary is a minimal value of the angular momentum of particles
with $E=m$ capable to achieve ergosphere falling from infinity.
    That is why collisions with $J_2 \to -\infty$ do not occur for particles
following from infinity.
    But if the particle came to ergosphere and there in the result of
interactions with other particles is getting large negative values of the
angular momentum projection (no need for getting high energies!)
then its subsequent collision with the particle falling on the black hole
leads to high energy in the centre of mass frame.

    Getting superhigh energies for collision of usual particles (i.e. protons)
in such mechanism occur however physically nonrealistic.
    Really from~(\ref{KerrJB}) the value of angular momentum necessary for
getting the collision energy $E_{\rm c.m.}$ has the order
    \begin{equation}
J_2 \approx - \frac{a E_{\rm c.m.}^{\,2}}{2 E_1}.
\label{KerrEMR}
\end{equation}
    So from~(\ref{KerrEM}) absolute value of the angular momentum $J_2$
must acquire the order $ E_{\rm c.m.}^{\,2} / (m_1 m_2)$ relative to the
maximal value of the angular momentum of the particle incoming to ergosphere
from infinity.
    For example if $E_1=E_2 = m_p$ (the proton mass) then $|J_2|$ must
increase with a factor $10^{18}$ for $ E_{\rm c.m.} = 10^9 m_p$.
    To get this one must have very large number of collisions with getting
additional negative angular momentum in each collision.

    However the situation is different for supermassive particles.
    In~\cite{GribPavlov2002(IJMPD)}--\cite{GribPavlov2008c} we discussed
the hypothesis that dark matter contains stable superheavy neutral particles
with mass of the Grand Unification scale created by gravitation in the end
of the inflation era.
    These particles are nonstable for energies of interaction of the order
of Grand Unification and decay on particles of visual matter but are
stable at low energies.
    But in ergosphere of the rotating black holes such particles due to
getting large relative velocities can increase their energy from $2m$ to
values of $3m$ and larger so that the mechanism considered in our paper
can lead to their decays as it was in the early universe.
    The number of intermediate collisions for them is not very large
(of the order of 10).

\vspace{9mm}
%%%% *****************************************************************
{\centering \section{Properties of movement of particles with negative energy
in ergosphere}
\label{secErgo}}

    From~(\ref{geodKerr3}), (\ref{geodTh}), (\ref{JmErg}) one can see that the
Carter's constant $Q \ge 0$ for particles with negative energy in ergosphere.
    One can have $Q=0$ for $E \le0$ only in case of movement in
equatorial plane.
    The value $Q=0$ is necessary for equatorial movement, but for the positive
energy $E$ with $Q=0$ the non-equatorial movement can take place.

    Let's consider other specifics for movements of the negative
energy particles.
    Define the effective potential by the formula
    \begin{equation} \label{Leff}
V_{\rm eff} = - \frac{R}{2 \rho^4}.
\end{equation}
    Then due to~(\ref{geodKerr3}),
    \begin{equation} \label{LeffUR}
\frac{1}{2} \left( \frac{d r}{d \lambda} \right)^{\!2} + V_{\rm eff}=0
\end{equation}
    and so
    \begin{equation} \label{LeffUR2}
\frac{d^2 r}{d \lambda^2} = - \frac{d V_{\rm eff}}{d r} .
\end{equation}

    Inside the event horizon up to Cauchy horizon
from~(\ref{geodKerr3}), (\ref{geodR}), (\ref{geodTh}), (\ref{JgEH})
one has $V_{\rm eff} <0$ for any falling particles.
    So any particle intersecting the event horizon must achieve the
Cauchy horizon.
    After going through the Cauchy horizon the particle can achieve
singularity.
    The necessary condition for this is that Carter constant $Q \le 0$
(see~(\ref{geodKerr3})--(\ref{geodTh})).
    For particles with negative energy in ergosphere this is true only for
movement in equatorial plane, i.e. $Q=0$.
    Further, one consider the case of equatorial movement $(\theta = \pi/2)$.

    Let us show that for particles with nonzero and zero masses (photons) with
negative relative to infinity energy there are no orbits inside ergosphere
in equatorial plane with constant $r$ or with $r$ changing for all geodesic inside
the interval $r_1 \ge r \ge r_H$~\cite{GribPavlov2013a}.

    The necessary condition of existence of orbits with constant $r$ is
    \begin{equation} \label{LeffCucl}
V_{\rm eff}=0, \ \ \ \ \frac{d V_{\rm eff}}{d r} =0\,.
\end{equation}
    To prove our statement it is sufficient to show that for
$dt/d \lambda >0$ and
    \begin{equation} \label{LeffdVefg0}
E<0, \ \ \ r > r_H, \ \ \ V_{\rm eff}(r)=0 \ \ \Rightarrow
\ \ V^{\, \prime}_{\rm eff}(r) > 0 .
\end{equation}

    For $V_{\rm eff}(r)=0$ the derivative of the effective potential
can be written as
    \begin{equation} \label{Leffder}
V_{\rm eff}^{\, \prime}(r) = \frac{1}{2 \rho^4}
\left( 2 (r -M) \frac{P^2}{\Delta} + 2 m^2 r \Delta - 4 r E P \right).
\end{equation}
    So in order to prove our statement it is sufficient to prove that
for the negative energy of the particle in ergosphere one has $P >0$.

    From the condition of movement ``forward in time'' and~(\ref{geodKerr1})
one obtains
    \begin{equation} \label{geodKett}
P \ge  \frac{ a \left( a E \sin^2 \! \theta - J \right) \Delta}
{r^2 + a^2}.
\end{equation}
    For particles in ergosphere with $E<0$ and $\theta=\pi/2$, $ r<2$
from~(\ref{lHmdd}), (\ref{JmErg}) one has
    \begin{equation}
a E  - J \ge a E  - J_{-} \ge \frac{a E }{ r^2 -2 r M} >0.
\label{LPerg}
\end{equation}
    That is why $P>0$ and our statement is proved.

    So there are no circular orbits for Penrose trajectories
in Kerr's black holes.
    The permitted zone for such particles in ergosphere can have only
upper boundary.

    Note that one gets absence of orbits with constant $r=r_H$ on the horizon
of the nonextremal black holes $a<M$ from the fact that for $V_{\rm eff}=0$
and $\Theta \ge 0$ one  has $J=2 r_H /M$ and
    \begin{eqnarray}
V^{\, \prime}_{\rm eff}(r_H)=
\frac{(r_H - M) (m^2 r_H^2 + (J-a E)^2 + Q}{\rho^4} \ge
\nonumber \\
\frac{(r_H-M)}{\rho^4 a^2} ( m^2 \rho^2 a^2 +
(r_H^4 - a^4 \cos^2 \! \theta ) E^2 +
4 r^2_H M^2 \cot^2 \! \theta ) > 0 .
\label{VderH}
\end{eqnarray}
    For extremal black holes $a=M$ one can see from~(\ref{VderH})
$V^{\, \prime}_{\rm eff}(r_H)=0$ for $\theta= \pi/2$.
    However circular orbits for $E\ne 0$ are also absent in this
case as it is shown in~\cite{HaradaKimura10}.

\vspace{9mm}
%%%% *****************************************************************
{\centering \section{The time of movement of particles with negative
energy in the ergosphere}
\label{secErttau}}

    Let us analyze the problem of the time of movement for particles with
negative energy in ergosphere.
    As it was shown in the previous section the geodesic with negative energy
in ergosphere begins from $r=r_H$, then achieves the upper point of the
trajectory $r_b$ and after it falls to horizon.
    So the proper time interval of movement of the particle along all geodesic
in ergosphere is defined by the integral
    \begin{equation} \label{VsIntdlbO}
\Delta \lambda = 2 \int \limits_{r_H}^{r_b} \frac{ d r }{|d r /d \lambda |}
= 2 \int \limits_{r_H}^{r_b} \frac{d r}{\sqrt{- 2 V_{\rm eff}(r)}} .
\end{equation}
    The factor 2 before the integral is due to taking into account the fact
that the proper time of movement along geodesic up from some value of
the radial coordinate $r$ is equal to the time of falling down to the same
value of $r$.

    In the vicinity of the upper point $r_b$
on the trajectory of the particle with negative energy one has
from (\ref{LeffUR}) and $V_{\rm eff}(r_b)=0$ that
    \begin{equation} \label{VsdrdlrH}
\left| \frac{d r}{d \lambda} \right| = \sqrt{- 2 V_{\rm eff}} \approx
\sqrt{2 (r_b - r) V^{\, \prime}_{\rm eff}(r_b)}.
\end{equation}
    As it was shown in the preceding part for the boundary point of the
permitted zone $V^{\, \prime}_{\rm eff}(r_b)>0$, so the integral
    \begin{equation} \label{VsIntdlb}
\int \frac{ d r }{|d r /d \lambda |} \sim
\int \frac{d r}{\sqrt{2 (r_b - r) V^{\, \prime}_{\rm eff}(r_b)}}
\end{equation}
    is convergent and the proper time of the lifting to the upper point
(falling from the upper point) of the trajectory in the vicinity of this
point is finite.

    Due to the fact that permitted zones for particles with negative energies
in ergosphere can have only upper boundary there are no zeros
for $dr / d \lambda$ in the other points of the trajectory.
    So the integral~(\ref{VsIntdlbO}) is convergent and the proper time of
movement along geodesic in the ergosphere for the particle with the negative
energy is finite.

    Note that the coordinate time of movement to the horizon is infinite.
    For equatorial movement the evaluations of the divergence of the
integral for coordinate time were given in~\cite{GribPavlov2013a}.

    Finiteness of the proper time for movement of particles with negative
energy in ergosphere of the black hole leads to the problem of the
origination and termination of such trajectories.
    As we said in the Introduction these lines cannot arrive to ergosphere
from the region outside of the ergosphere.
    So they must originate and terminate inside the gravitational radius.
    This means that they originate as ``white hole'' geodesics originating
inside the horizon.

    Note that similar situation takes place for some geodesics of particles
with positive energy in Schwarzschild metric (see the text book~\cite{LL_II}).
    The geodesic completeness leads to the necessity of taking into account
``white hole'' geodesics originating in the past singularity of the eternal
black hole for radial geodesics with specific energy $E/m<1$
arising from the region inside the gravitational radius.
    However for Penrose geodesics we show that all such geodesics
in ergosphere of the Kerr's black hole have such behaviour.

%%%%%%%%%%%%%%%%%%%%%%%%%%%%%%%%%%%%%%%%%%%%%%%%%%

    From~(\ref{geodKerr3})--(\ref{ThB0}), (\ref{JgEH}) one can see that all
light like particles (photons) with negative energy falling in equatorial
plane from ergosphere achieve singularity.
    Massive particles also achieve singularity for example if $E=-m$
or the angular momentum is such that
    \begin{equation} \label{FalSin}
J \le J_H \biggl( 1 + \frac{a}{2M} \sqrt{1 + \frac{m^2}{E^2} } \, \biggr).
\end{equation}
    The proof of all these results is the same for
$d r / d \lambda >0$ and $d r / d \lambda < 0$.
    That is why the same conditions are valid for ``white hole''geodesics
originating in Kerr's singularity, arriving to ergosphere and then going back
inside the gravitational radius.
    Particles moving in equatorial plane do not achieve singularity
if for example
    \begin{equation} \label{FalNS}
\frac{|E|}{m} \ll \frac{r_C}{M} , \ \ \frac{|J|}{m M} \ll \frac{r_C}{M} .
\end{equation}
    Then after achieving some minimal values of the radial coordinate
the particle can turn it's movement in the direction of larger $r$ and come
back to ergosphere along the white hole geodesics.

    Let's come back to the question on the energy of collisions for
particle falling onto black hole and  the particle with negative energy
moving in ergosphere.
    As it was shown previously (Eqs.~(\ref{JmErg}), (\ref{JgEHf}))
particles with large in absolute value negative angular momentum can achieve
the region close to the upper frontier of ergosphere.
    The energy of collision of such particle with the ordinary particle
will be very large due to~(\ref{KerrJB}) at any point of the ergosphere.

    The existence of particles moving from the gravitational radius in the
direction of larger $r$ along white hole geodesic can give us
the new opportunity for collisions independent of angular momentum
with non-limited energy near black holes.
    As we can see from~(\ref{KerrL1L2}) the difference between energy
of collisions with particle moving to increasing $r$  $ (\sigma_{2r} =1)$
and decreasing $r$  $ (\sigma_{2r} =-1)$ is
    \begin{equation} \label{DDD}
\Delta E_{c.m}^2 = \frac{4 \sqrt{R_1} \sqrt{R_2}} {\rho^2 \Delta} .
\end{equation}
    This difference is equal to zero for top point of trajectory,
when $R_2 =0$.
    But for non-critical particles $(J \ne J_H)$ the difference is infinite
large for collision on the horizon $(r \to r_H)$.
    So from~(\ref{KerrL1L2}) one has for such collisions
$(\sigma_{1r} \sigma_{2r} =-1)$ for $r\to r_H$: \,
$P_1 P_2 - \sigma_{1r} \sqrt{R_1} \sigma_{2r} \sqrt{R_2} > 0$, \, $\Delta \to 0$,
    \begin{equation} \label{ss12m}
E_{c.m} \sim \frac{ \sqrt{ 2 (P_1 P_2 - \sigma_{1r} \sqrt{R_1} \sigma_{2r} \sqrt{R_2})}}
{\rho \sqrt{\Delta}} \approx
\frac{2 a}{ \sqrt{r-r_H} } \sqrt{
\frac{(J_{1H} - J_1 ) (J_{2H} - J_2 ) }
{(r_H^2 + a^2 \cos^2 \theta )(r_H-r_C)}} \to +\infty .
\end{equation}
    The analogue of the result~(\ref{ss12m}) is valid not only for Kerr's black
holes but also for Schwarzschild black holes
    \begin{equation} \label{SHw}
a=0, \ \ \sigma_1 \sigma_2 =-1, \ \ \
E_{c.m} \sim 2 \sqrt{ \frac{E_1 E_2}{ 1 - (r_H/ r)}} \to +\infty, \ \
r \to r_H=2M.
\end{equation}
    As it is known from the textbook~\cite{HawkingEllis}
for the Schwarzschild case geodesic completeness leads to existence of
``white hole'' geodesics.
    Note that the energies of particles $E_1, E_2$ can not be negative
outside event horizon of Schwarzschild black holes~\cite{GribPavlov2010NE}.
    One can see from~(\ref{Relsk03}) that the effect of the large energy
is explained by the large Lorentz factor, so that $v_{\rm rel} \sim 1$
on the horizon.

    The formulas~(\ref{ss12m}), (\ref{SHw}) can be interpreted as nonstability
of the configuration with ``white hole'' geodesics and therefore instability of
eternal black holes.
    Evaluation of the role of the gravitational wave radiation for large
values of the energy in the centre of mass frame is needed to get the final
answer about the physical sense of the obtained results.

\vspace{7mm}
%%%% *****************************************************************
{\centering \section{Conclusion}
\label{Conclusion}}

1) Geodesics with negative energy in equatorial plane
originate inside the gravitational radius of the rotating black hole
are ``white hole'' geodesics!
    This is similar to the case of Schwarzschild eternal black hole for which
it was shown in~\cite{GribPavlov2010NE} that negative energy
trajectories arise in the white hole past singularity.
    However differently from the eternal Schwarzschild black hole when there
are two different space like singular surfaces --- the black hole and white
hole --- here we have one Kerr's time like singular surface on which some
geodesics originate and some terminate.

2)
One can get information about the interior of the gravitational radius if
some particles move along these geodesics!
    So there is no cosmic censorship for such rotating black holes if one
understands cosmic censorship as impossibility to get information from
the region inside the gravitational radius.
    The cosmonaut can get direct information about the interior of the
gravitational radius only inside the ergosphere.
However if one considers interaction of negative energy particles radiated by
the ``black-white'' hole to ergosphere with ordinary positive energy particles
escaping the ergosphere this information can be obtained by any external observer.
    The electromagnetic interaction of the negative energy photons with usual
matter can lead to some new physical process of ``annihilation'' of this
matter inside ergosphere.
    The energy in the centre of mass frame of the collision of ordinary
positive energy particle with the particle with negative energy is growing
without limit on the horizon.

3)
The mass of the ``black-white'' hole can grow due to the radiation of negative
energy particles to ergosphere where these particles interact with
positive energy particles.

    Surely all results in this paper correspond to exact Kerr's solution.
    For physically realized rotating black holes as the result of the star
collapse due to nonstability of interior solution inside the gravitational
radius especially inside the Cauchy horizon (see~\cite{GribPavlov2008UFN})
the results can be different and the special research is needed.

\vspace{0mm}
%%%% *****************************************************************
 {\bf Acknowledgments.}
    The research is supported by the grant from The John Templeton Foundation.

\vspace{5mm}
%%%% ****************************************************************

\end{document}